\documentstyle[11pt,twocolumn]{article}
\begin{document}
\twocolumn[\begin{center}

{\large REPORT No.~~1681/PH}\\

\vskip1cm

{ \Large  PHYSICS AT LOW {$x~ ^*$}}\\

\vskip1cm

 J. Kwieci\'nski\\

\vskip1.5cm

 Henryk Niewodnicza\'nski Institute of Nuclear Physics\\
\smallskip

Department of Theoretical Physics\\
\smallskip

ul. Radzikowskiego 152,~~ 31-342~ Krak\'ow, Poland\\
\end{center}

\vskip5truecm
\noindent  -----------------------------------

\noindent $^*$  Review talk presented at the Workshop
"QCD'94", 7-13 July 1994, Montpellier, France]

\pagebreak

\twocolumn[\begin{center}
\vskip1cm
{\bf PHYSICS AT LOW $x$}\\
\vskip0.5cm

J. Kwieci\'nski\\

{\it Department of Theoretical Physics}\\
{\it H. Niewodnicza\'nski Institute of Nuclear Physics}\\
{\it ul. Radzikowskiego 152, 31-342 Krak\'ow, Poland}\\
\end{center}
\vskip0.5cm

{\small The QCD expectations concerning the deep inelastic lepton - hadron
scattering
 at low $x$ and their phenomenological implications for HERA are summarised.
Theoretical predictions for the structure function $F_2(x,Q^2)$ based
on the leading log$1/x$ resummation are presented and compared with the
results obtained from the Altarelli-Parisi equations. Theoretical
predictions are confronted with the recent data from HERA. The role of
studying the final states in deep inelastic scattering for revealing the
dynamics at low $x$ is emphasised and some dedicated measurements
like  deep inelastic plus jet events, transverse
energy flow and dijet production in deep inelastic scattering are
discussed.

\vskip1cm

\phantom{of the opera} }]

 In  this talk we shall give an overview of the QCD expectations concerning
the small $x$ behaviour of deep inelastic lepton-hadron scattering
and will discuss their phenomenological implications for HERA.
Besides discussing the deep inelastic structure functions  we shall also
consider specific measurements like deep inelastic scattering
accompanied by the energetic jet, transverse energy flow and dijet
production in deep inelastic scattering
which should test the
QCD predictions at small $x$  more directly and unambiguosly.

The dominant role at small $x$ is played by the gluons and so at first
we consider the small $x$ limit of the gluon distribution. The
gluons are not, of course, directly probed in deep inelastic
lepton
scattering, yet they contribute indirectly to this process
through the $g \rightarrow q\bar q$ transitions.

The relevant
framework for calculating the parton distributions in perturbative
QCD in the small $x$ limit is the leading log$(1/x)$ (LL$1/x$) approximation.
It corresponds to the sum
of those terms which contain the maximal power
of $ln(1/x)$ at each order of the perturbative expansion.  The basic quantity
in this approximation is the unintegrated gluon distribution $f(x,k^{2})$
which is related in the following way to the conventional (scale dependent)
gluon distribution $g(x,Q^{2})$:
\begin{equation}
xg(x,Q^{2})=\int_{0}^{Q^{2}}{dk^{2}\over k^{2}}f(x,k^{2})
\label{xg}
\end{equation}
In the LL$1/x$
approximation the unintegrated gluon distribution satisfies the following
equation \cite{bfkl}-\cite{kk}:

$$  - x{\partial f(x,k^{2})\over \partial x}=$$
$${3\alpha_s(k^{2})\over \pi}k^{2}\int_{k_{0}^{2}}^\infty
{dk'^{2}\over k'^{2}}\Big[
{f(x,k'^{2})-f(x,k^{2})\over\vert
k'^{2} - k^{2}\vert} $$
$$+ {f(x,k^{2})\over \sqrt{4k'^{4} + k^{4}}}\Big]$$
\begin{equation}
\equiv K_{L}\otimes f
\label{bl}
\end{equation}
which is called in the literature the Balitzkij-Fadin-Kuraev-Lipatov
(BFKL) equation.

It corresponds to the sum of ladder diagrams
but unlike
the leading log$Q^{2}$ approximation which gives the Altarelli-Parisi
evolution equations \cite{ap} the transverse momenta of the gluons
are not ordered along the chain.
The kernel $K_{L}$ of the eq.(\ref{bl})
contains both the real gluon emission term
as well as the virtual corrections.  The former corresponds to the term
proportional to $f(x,k'^{2})$ while the latter correspond
to terms proportional to
$f(x,k^{2})$ in the integrand on the rhs. of the eq.(\ref{bl}).
The virtual corrections
correspond
to the "gluon reggeisation" \cite{bfkl,glr} (or to the "non-Sudakov"
form factor
\cite{cia,mar}).
The variable(s) $k^{2} (k'^{2})$ denote the transverse
momenta squared of the gluons along the ladder. The parameter $k^{2}_{0}$ is
the
infrared cut-off which is necessary
if the running coupling constant effects are taken into account.
More general treatment of the infrared region in the BFKL
equation than simply imposing the lower limit cut-off $k_0^2$
has been discussed in \cite{fh}-\cite{hr}.

When the running coupling effects are neglected (i.e. when one sets
$\alpha{_s}(k^{2})=\bar \alpha{_s}$) and when $k^{2}_{0}=0$ then
the equation (\ref{bl}) can be solved analytically and the leading small $x$
behaviour of its solution is  given by the following formula:
$$f(x,k^{2})\sim $$
\begin{equation}
{x^{-\lambda}\over[ln(1/x)]^{1/2}}
(k^{2})^{1/2}exp\left (-{ln(k^2/\bar k^2)\over
2\lambda"ln(1/x)}\right )
\label{dif}
\end{equation}
with
\begin{equation}
\lambda = {12ln(2)\over \pi}\bar \alpha_{s}
\label{lam}
\end{equation}
\begin{equation}
\lambda"= {3\bar \alpha_{s}\over \pi}28\zeta (3)
\label{lamb}
\end{equation}
where the Riemann zeta function $\zeta (3) =1.202$.  The
parameter $\bar k$ is of nonperturbative origin and is fixed by
the boundary condition for $f(x,k^2)$ at $x=x_0$ \cite{akms1}.

The solution (\ref{dif}) of the BFKL equation follows from the solution of
the corresponding equation for the moment $\bar f(n,k^2)$
\begin{equation}
\bar f(n,k^2) = \int_{0}^{1} dx x^{n-2} f(n,k^2)
\label{momf}
\end{equation}
of the
distribution $f(x,k^2)$

$$\bar f(n,k^2)=\bar f^0(n,k^2)+$$
$${3\bar\alpha_s\over \pi (n-1)}
k^{2}\int_{0}^\infty {dk'^
{2}\over k'^{2}}\Big[
{\bar f(n,k'^{2})-f(n,k^{2})\over\vert
k'^{2} - k^{2}\vert}$$
\begin{equation}
+ {f(n,k^{2})\over \sqrt{4k'^{4} + k^{4}}}\Big]
\label{blmom}
\end{equation}
where $\bar f^0(n,k^2)$ is the moment of the suitably defined inhomogeneous
term.  This equation is diagonalised by the Mellin transform and its solution
for the Mellin transform $\tilde f(n,\gamma)$ of the moment $\bar f(n,k^2)$
is:
\begin{equation}
\tilde f(n,\gamma)={\tilde f^0(n,\gamma)\over 1-
{3\bar \alpha_s\over \pi (n-1)
}\tilde K_L(\gamma)}
\label{blmel}
\end {equation}
where
\begin{equation}
\tilde K_L(\gamma)=2\Psi(1) - \Psi(\gamma) - \Psi(1-\gamma)
\label{klmel}
\end {equation}
is the Mellin
transform of the kernel of the equation (\ref{blmom}).  The function $\Psi(z)$
is the logarithmic derivative of the Euler $\Gamma$ function. The Mellin
transform $\tilde f(n,\gamma)$ is defined as below:
\begin{equation}
\tilde f(n,\gamma)=\int_{0}^{\infty}dk^2(k^2)^{-\gamma-1}f(n,k^2)
\label{mel}
\end{equation}

The poles of $\tilde f(n,\gamma)$ in the $\gamma$ plane define the gluon
anomalous dimensions $\bar \gamma(n,\bar \alpha_s)$.  The leading
twist anomalous dimension $\bar \gamma^{LT}(n,\bar \alpha_s)$
controls the $k^2$ dependence of
$\bar f(n,k^2)$ at large $k^2$
\begin{equation}
\bar f(n,k^2) \sim (k^2)^{\bar \gamma^{LT}(n,\bar \alpha_s)}
\label{lkf}
\end{equation}

    It follows from
the eq. (\ref{blmel}) that the anomalous dimensions are the solutions of
the following
equation:
\begin{equation}
{3\bar \alpha_s\over \pi (n-1)} \tilde K_L(\bar \gamma(n,\bar \alpha_s))
=1
\label{adim}
\end{equation}
The solution of this equation allows to obtain the gluon
anomalous dimensions
as a power series of ${3\bar \alpha_s\over \pi (n-1)}$ \cite{jar,ekl}.
For the leading twist anomalous dimension this power series corresponds to
the leading log$1/z$ expansion of the splitting  function
$P_{gg}(z,\alpha_s)$
which appears in the evolution equation for the gluon distribution.
The leading twist quark anomalous dimension which includes the
resummation
of the  powers of ${3\alpha_s\over \pi (n-1)}$ has recently been
discussed in the ref. \cite{qad}.

The exponent $\lambda$ controlling the
small $x$ behaviour of $f(x,k^2)$ (cf. equations (\ref{dif}, \ref{lam}))
is
\begin{equation}
\lambda={3\bar \alpha_s\over \pi}\tilde K_L(1/2)
\label{lam1}
\end{equation}
The leading twist anomalous dimension has a branch point singularity
at $n=1+\lambda$. We also have (cf. eqs. (\ref{adim}, \ref{lam1})):
\begin{equation}
\bar \gamma^{LT}(n=1+\lambda,\bar \alpha_s) = {1\over 2}
\label{adims}
\end{equation}

The following properties of the solution of the BFKL equation
 summarised in the formula (\ref{dif}) should be noted:

(1) It exhibits the Regge type  $x^{-\lambda}$  increase
with
decreasing $x$ where the
 exponent $\lambda$ given by eq. (4) can have potentially large
magnitude $\simeq 1/2$ or so. The quantity $1+\lambda$
is equal to the intercept of the so called BFKL
Pomeron which corresponds to the hard QCD interactions.

Its potentially large magnitude ($\simeq$ 1.5) should be contrasted
with the intercept  $\alpha_{soft} \approx 1.08$ of the
(effective) "soft" Pomeron
which has been determined from the phenomenological analysis of the high
energy behaviour of hadronic and photoproduction total
cross-sections \cite{dl1}.

(2) It exhibits the $(k^2)^{1/2}$ growth with increasing
$k^2$ modulated by the Gaussian distribution in $ln(k^2)$ with its width
increasing as $ln^{1/2}(1/x)$ with
decreasing $x$.  The Gaussian factor reflects the diffusion
pattern of the  BFKL equation \cite{bfkl,akms1,blot}.  The increase of the
function $f(x,k^2)$ as $(k^2)^{1/2}$ is due to the fact that the
leading twist anomalous dimension is equal to 1/2 at the BFKL singularity
(cf. eq. (\ref{adims})).
This shift of the anomalous dimension is the result
of the (infinite) $LL(1/x)$ resummation
and should be contrasted with the anomalous dimension calculated
perturbatively retaining only
finite number of terms \cite{ekl}.

If the running coupling effects are taken into account then the leading
small $x$ behaviour is:
\begin{equation}
 f(x,k^{2}) \sim x^{-\bar\lambda} \label{smxr}
 \end{equation}
The exponent $\bar\lambda$ has to be now calculated numerically
and is dependent
upon the infrared cut-off $k^{2}_{0}$ \cite{fh,hr,kms1,kms2}.
More recent discussion of the BFKL equation is given in \cite{cl,muel1}.

The validity of the BFKL equation is, in principle, restricted to the
region where $\alpha_s ln(1/x) \sim O(1)$ and $\alpha_s ln(Q^{2}/Q_{0}^{2})
<< 1$ where $Q_{0}^{2}$ is some moderately large
scale ($Q_{0}^{2}>>\Lambda^{2}$
).  Possible extension of the BFKL equation beyond this region is provided
by the Marchesini equation \cite{mar} which  treats both (large)
logarithms
$ln(1/x)$ and $ln(Q^{2}/Q_{0}^{2})$ on equal
footing. In the region of large values of $x$ this equation becomes
equivalent to the Altarelli-Parisi evolution equations.

For $xg(x,Q^{2})$ growing as $x^{-\lambda}$ the
transverse area $\tilde S(x,Q^{2})$ occupied by the gluons
\begin{equation}
\tilde S(x,Q^{2})
= const~xg(x,Q^{2}){\alpha_s(Q^{2})\over Q^{2}}\label{st}
\end{equation}
 can, for  sufficiently small value
of $x$ and for fixed $Q^{2}$ become comparable to the transverse area
of a hadron $S_{H}=\pi R_{H}^{2}$
where $R_{H}$ is the hadronic radius.  When this happens (and in fact before
this happens),  the gluons can no longer be treated as free partons
and their interaction leads to screening (or shadowing) effects \cite
{glr,kk,mq}.
The main effect of shadowing is to tame the indefinite increase
of parton distributions.  One finds instead that at sufficiently small
values of $x$ and/or $Q^{2}$ the gluon distributions approach the so-called
saturation limit  $xg_{sat}(x,Q^{2})$ \cite{glr,kk}
\begin{equation}
xg_{sat}(x,Q^{2})={const\over\alpha_s(Q^{2})}R_{H}^{2}Q^{2} \label{satl}
\end{equation}
In some models \cite{muel2} the saturation limit contains some remnant
weak $x$ dependence.  The most dramatic effect is the linear scaling violation
exhibited by $xg_{sat}(x,Q^{2})$.

If one assumes that the gluons are not distributed uniformly within a hadron
but are  concentrated around the "hot-spots" \cite{lr1,muel2} having their
 radius $R_{h.s.}$ much smaller than the hadronic radius  $R_{H}$
then the shadowing
effects are expected to be stronger.  The saturation limit $xg_{sat}(x,Q^{2})$
is then controlled by the radius $R_{h.s.}$ and not by $R_{H}$.

The shadowing effects modify
the BFKL equation by the non-linear terms \cite{glr,kk,kms1,agr}:

$$-x{\partial f(x,k^{2})\over \partial x}=$$
\begin{equation}
K_{L}\otimes f
 -{81\alpha^{2}_s(k^{2})\over 16R^{2}k^{2}}[xg(x,k^{2})]^{2}
 \label{glrs}
\end{equation}
The equation (\ref{glrs}) is called in the literature the Gribov, Levin,
Ryskin (GLR) equation.

The second term in the right hand side of the eq.(\ref{glrs}) which is
quadratic
in the gluon distribution describes the shadowing effects.
They correspond to the QCD diagrams where two gluonic ladders
merge into one.
The parameter $R$ describes the size of the
region within
which the gluons are concentrated.

The GLR equation has been derived assuming that  the gluons
within a hadron are uncorelated and that the interaction between
the gluonic ladders can be neglected. These assumptions have recently been
shown to be unjustified \cite{bart1}.  The  interaction
between gluonic ladders can be
 approximmately accounted for by introducing the appropriate enhancement
 factor in the nonlinear term in the eq. (\ref{glrs}) \cite{br}.

It turns out that in the region of $x$ and $Q^{2}$ which may
be relevant for HERA the shadowing effects are still relatively weak
at least for $R \sim 5 GeV^{-1}$ \cite{kms1}.  The possibility of detecting
the nonlinear shadowing terms in HERA has recently been discused
in detail in the ref. \cite{gkra}.

In Fig.1 we show the regions in the (ln$1/x$, ln($Q^2/\Lambda^2$))
plane
where various dynamical effects and approximations
discussed above are expected to play the dominant role.

We shall now discuss quantitative predictions for the structure function
$F_2(x,Q^2)$  at small $x$ which will be directly
based on the solution of the BFKL equation \cite{akms1,akms2,aja}.  (For
the first attempt to determine the structure function $F_2(x,Q^2)$ from
the solution of the BFKL equation see the ref. \cite{dsk}).
The relevant diagrams are shown in Fig.2 and their contribution to the
structure
function $F_2(x,Q^2)$ can be written in the following factorisable form:
$$ F_{2}(x,Q^2) = $$
\begin{equation}
\int_{x}^{1} {dx'\over x'} \int {dk_T^2\over k_T^4}f({x\over x'},k_T^2)
F^{0}_{2}(x',k_T^2,Q^2)
\label{ktf}
\end{equation}
where $x/x'$ is the longitudinal momentum fraction carried by the gluon
which couples to the $q \bar q$ pair.  Similar representation can be
written for the longitudinal structure function $F_{L}(x,Q^2)$.
The function $F^0_{2}$ which
denotes the quark box contribution to the photon-gluon subprocess
shown in Fig.1 is given in ref. \cite{akms2}. The function $f(x/x',k_T^2)$
is the unintegrated gluon distribution corresponding to the
sum of ladder diagrams  in the lower part of the diagrams shown in Fig.2.
In the leading log$(x'/x)$ approximation this distribution is given as
the solution of the BFKL equation (\ref{bl}). The $(x/x')^{-\lambda}$
behaviour of the unintegrated gluon distribution $f$ (see the equations
(\ref{dif}, \ref{smxr})) generates the singular $x^{-\lambda}$ behaviour
of the structure function $F_2(x,Q^2)$.  We may also study
the effects of shadowing in the structure functions by using the function
$f$ which is the solution of the GLR equation (\ref{glrs}).

The $Q^2$ dependence of the structure function $F_2(x,Q^2)$  reflects
the $k_T^2$ dependence of the function $f(x,k_T^2)$ \cite{aja}.
Thus, for instance, the
increase of the unintegrated  gluon distribution as $(k_T^2)^{1/2}$
with increasing $k_T^2$ (see the eq. (\ref{dif})) implies increase of
the structure function
$F_2(x,Q^2)$ as $(Q^2)^{1/2}$ with increasing $Q^2$.

The formula (\ref{ktf}) is an example of the $k_{T}$ factorisation
theorem which is the basic tool for calculating the
observable quantities (i.e. in this case the
structure functions) at small $x$ \cite{kt}. Its connection
with the more conventional collinear factorisation has recently
been discussed in detail in the ref. \cite{kt2}.

In Fig.3  we summarise the predictions for the structure
function  $F_2(x,Q^2)$ based on the formula (\ref{ktf})
\cite{akms1,akms2}  (the curves marked as AKMS)
and confront them with the recent
data from HERA \cite{hera1,ubass,hjung}. We also compare these
predictions with structure
functions which were obtained within the NLO Altarelli-Parisi formalism.
Let us recall that in the latter case the small $x$ behaviour
depends upon the phenomenological input distributions at the
reference scale and is therefore not constrained by a theory.
One can mimick the BFKL behaviour at small $x$ by assuming the
$x^{-\lambda}$ type extrapolation of the input parton
distributions
as it has been done in the refs. \cite{mrs1,mrs2,mrs3}
(the curves MRS(A), \cite{mrs3}).
In the  next-to-leading approximation one neglects,
of course,
the effects of the leading log$1/x$ resummation in the splitting
(and coefficient)
functions which are automatically taken care of in the calculation based
on the $k_T$ factorisation with the function $f$ obtained from the solution
of the BFKL equation.  Systematic study of those effects within the
Altarelli -Parisi formalism has however shown that their role is
relatively small
provided the starting distributions are sufficiently singular at small $x$
\cite{ekl}. This result explains possible origin of similarity between
the AKMS and MRS curves.

In the plots in Fig.3 which correspond to $Q^2=$ 15 GeV$^2$ and
to $Q^2=$ 30 GeV$^2$
we also show  predictions
obtained from the  "dynamical model" of parton distributions \cite{grv}.  In
this model the parton distributions are generated radiatively
from
the valence-like input at the very low reference scale $Q_0^2
\simeq 0.25 GeV^2$.  The strong increase of $F_2(x,Q^2)$
with decreasing $x$ at the HERA range (i.e. for $Q^2 \sim 10
GeV^2$)  comes now from the relatively large
evolution length $\xi (Q^2,Q^2_0)$
\begin{equation}
\xi (Q^2,Q^2_0)=\int_{Q_0^2}^{Q^2}{dq^2\over
q^2}{3\alpha_s(q^2)\over \pi} \label{ksi}
\end{equation}
since in this case $F_2(x,Q^2)$ behaves at small $x$ as below:
\begin{equation}
F_2(x,Q^2) \sim exp\left (2 \sqrt{\xi
 (Q^2,Q^2_0)ln(1/x)}\right ) \label{dlf2}
\end{equation}

The structure function $F_2(x,Q^2)$ which
is generated radiatively in the leading log$Q^2$  approximation
from the non-singular input distributions
at small $x$, exhibits
scaling in the two
variables $\sqrt{ln(1/x)lnlnQ^2}$ and $\sqrt{ln(1/x)/lnlnQ^2}$.
Exploration of this scaling property has recently been advocated as a
possible discriminator between the BFKL-motivated or purely radiatively
generated structure functions  and evidence has been
found for supporting the latter \cite{forte}.
Above scaling property  of structure functions may
however be modified if the leading log$(1/x)$ terms are included
in the splitting and coefficient functions. These terms play
significant role in the case of the non-singular starting
parton distributions \cite {ekl}.

We notice from Fig.3 that in the region of $x$ and $Q^2$ relevant for
HERA measurements the structure function $F_2(x,Q^2)$ which was
based on the solution of the BFKL equation is similar to that which was
obtained within the Altarelli-Parisi  evolution equations formalism in the
NLO approximation. The inclusive quantity like $F_2$ is not therefore the best
discriminator for
revealing the dynamical details at low $x$.

One may however hope that this can be provided by  studying
the structure of the final states at small $x$.  It is expected
in particular that absence
of transverse momentum ordering which is the characteristic
feature of the BFKL dynamics should reflect itself in the less
inclusive quantities than the structure function $F_2(x,Q^2)$.

The dedicated measurements of the low $x$ physics i.e.
the deep inelastic
plus jet events, transverse energy flow in deep inelastic scattering,
production of jets separated by the large rapidity gap and dijet
production in deep inelastic scattering are summarised in Fig.4,
\cite{admb}.

The deep inelastic lepton-hadron scattering containg a measured jet (Fig. 3a)
can provide a very clear test of the BFKL dynamics at small $x$
\cite{kms2,disj,kms3}.
The idea is to study deep inelastic ($x,Q^{2}$) events which contain
an identified jet ($x_{j},k^{2}_{Tj})$ where $x<<x_{j}$ and $Q^{2}\simeq
k^{2}_{Tj}$.   Since we choose events
with $Q^{2}\simeq k^{2}_{Tj}$ the QCD evolution (from $k^{2}_{Tj}$ to
$Q^{2}$) is neutralised and attention is focussed on the small $x$, or rather
small $x/x_{j}$ behaviour. The small $x/x_{j}$ behaviour of the jet production
is generated by the gluon radiation as shown in the diagram of Fig.4a .
Choosing the configuration $Q^{2}\simeq k^{2}_{Tj}$ we eliminate by definition
the gluon emission which corresponds to strongly ordered transverse momenta
i.e. that emission which is responsible for the QCD evolution.
The measurement of jet production in this configuration may therefore
test more directly the $(x/x_{j})^{-\lambda}$ behaviour which is generated
by BFKL equation where the transerse momenta are not ordered.

Let us now discuss the feasibility of using the deep inelastic events
which contain a meaured jet to identify the singular $z^{-\lambda}$ type of
behaviour at HERA [16,41,42].

One practical limitation is that jets can only be measured if they are emitted
at sufficiently large angles $(\theta_{j}>5^{0})$ to the proton beam
direction  in the HERA laboratory frame.
Large $x_{j}$ jets are only emitted at small $\theta_{j}$; for a given
$\theta_{j}$ we can reach larger $x_{j}$ by observing jets with larger
$k_{Tj}^{2}$ but with a depleted event rate.
In order to identify the BFKL $z^{-\lambda}$ behaviour we need deep
inelastic
+ jet events, with $k_{Tj}^{2}\simeq Q^{2}$, over an interval of
$z\equiv x/x_{j}$ which covers values of z as small as is experimentally
possible.  As a compromise we select the region $x_{j}>0.05$ and $x<2*10^{-3}$.

Fig.5 shows the
predicted $x$ dependence of the deep inelastic+jet cross-section relevant
for HERA \cite{kms2,kms3}. (The cross-section shown in Fig.5 has been
calculated
assuming the following constraints:
$x_{j}>0.05$ and $\theta_{j}>5^{0}$,
${1\over 2}Q^{2}<k_{Tj}^{2}<2Q^{2}$).
The continuous curves give the values of the cross-section when
the BFKL
effects are included.  These are to to be contrasted with the dashed curves
which show the values when the BFKL  effect is neglected, that is when just
the quark box approximation is used to evaluate the corresponding differential
tructure functions.  The steep rise of the
continuous curves with decreasing $x$ (i.e. decreasing $z\equiv x/x_{j}$)
reflects the $z^{-\lambda}$ BFKL effect generated by the
gluon radiation.

The recent H1 results concerning the deep
inelastic plus jet events which were reported at this
Workshop \cite{ubass} are consistent with the increase of the
cross-section with decreasing $x$ as predicted by the BFKL dynamics.

Conceptually similar process is that of the two-jet production
separated by a large rapidity gap $\Delta y$ in hadronic
collisions \cite{mnav,stir1} or in photoproduction \cite{stir2} as
illustrated in the Fig.4c.
Besides the characteristic $exp(\lambda \Delta y)$ dependence of the two-jet
cros-section one expects significant weakening of the azimuthal
back-to-back correlations of the two jets. This is the direct conseqence of
the absence of transverse momentum ordering along the gluonic  chain in the
diagram of Fig.4c. The experimental data on the dijet production
have recently been obtained at the Tevatron and were reported in this
Workshop \cite{d0}.

Another measurement which should be sensitive to the QCD
dynamics at small $x$ is that of the
transverse energy flow in deep inelastic lepton scattering \cite{brw,kmsg}.
The transverse energy flow in the central region away
from the current jet and the proton remnants can be calculated from
the following formula \cite{kmsg}:
$${\partial E_T\over \partial y}=$$
$${1\over F_2}\int dk_j^2 \vert k_j \vert
\int {d^2k_p\over \pi k_p^4}\int {d^2k_{\gamma}\over k_{\gamma}^4}
\left({3\alpha_s\over \pi}{k_p^2k_{\gamma}^2\over k_j^2}\right)$$
$${\cal F}_2(x/x_j,k_{\gamma}^2,Q^2)f(x_j,k_p^2)$$
\begin{equation}
\delta^{(2)}(k_j-k_{\gamma}-k_p)
\label{et}
\end{equation}
where the transverse momenta are defined in Fig.6a.  The function
${\cal F}_2$
describes the gluon radiation in the uper part of the diagram in Fig.6 a.
It satisfies the BFKL equation as does the unintegrated gluon distribution
$f$.
The variable $x_j$ is related to the rapidity $y(cm)$
in the virtual photon-proton
center-of-mass frame
\begin{equation}
y(cm)={1\over 2}ln\left({x_jQ^2\over xk_j^2}\right)
\label{ycm}
\end{equation}
The rapidity $y$ in the HERA frame is approximately related to $y(cm)$ by
a simple boost

\begin{equation}
y=y(cm) +  {1\over 2}ln\left({4xE_p^2\over Q^2}\right)
\end{equation}
since at small $x$ the two frames are approximately collinear.
The formula (\ref{et}) is only valid in a region where both $x_j$ and
$x/x_j$ are suffiiently small for the BFKL equation for $f(x_j,k^2_p)$
and for ${\cal F}_2(x/x_j,k^2_{\gamma},Q^2)$ to be valid (we assume
$x_j <10^{-2}$ and $x/x_j<10^{-1}$). In particular, for large $x_j$,
($x_j>10^{-2}$) the more appropriate way to calculate the energy flow is by
the use of the following formula:
$${\partial E_T\over \partial y}=$$
$${1\over F_2}\int {dk_j^2\over k_j^4} \vert k_j \vert
\left({3\alpha_s\over \pi}\right)
{\cal F}_2(x/x_j,k_{j}^2,Q^2)$$
\begin{equation}
\sum_af_a(x_j,k_j^2)
\label{eth}
\end{equation}
which follows from the strong ordering, $k_j^2>>k_p^2$,
at the gluon emission vertex (cf. Fig.6b) where the "effective" parton
combination $\sum_a f_a$ can be set equal to $g+{4\over 9} (q+\bar q)$.

The integral in the equation (\ref{et}) is weighted towards large values
of the tranverse momenta and so it is sensitive to the behaviour
of the functions
$f$ and $\cal F$ for large $k_p^2$ and $k_{\gamma}^2$ respectively.
In the central rapidity region away from the current jet and the
proton remnants the BFKL dynamics does predict substantial
amount of transverse energy which grows with decreasing $x$.

For fixed QCD coupling $\bar \alpha_s$ one gets the
$x^{-\epsilon}$ increase of transverse energy with decreasing
$x$ with $\epsilon=(3\bar \alpha_s/\pi)2 ln2$ \cite{kmsg}.  This increase
is closely related to the diffusion pattern of the solutions of the BFKL
equation for the functions $f$ and $\cal F$ (cf. eq. (\ref{dif})).
It should be noted that without the Gaussian factors which broaden with
decreasing $x_j$ or $x/x_j$ the integral defining the energy flow
would be divergent.

Numerical estimate  of the energy flow based on the BFKL equation with
running
coupling effects incorporated confirms the increase of the avarage $E_T$
with decreasing $x$ although this increase is weaker than for the
fixed coupling case. The
average value of $E_T$ per unit rapidity which is generated by
 the BFKL dynamics has been estimated to
be around 2 GeV. It is interesting to notice that the H1
collaboration has seen the excess of transverse energy in the
forward region \cite{ubass,eth1} in comparison to the expectations based on
the standard Monte Carlo models incorporating conventional QCD
cascades. In Fig.7 we confront the theoretical predictions
for the transverse energy flow based
on the BFKL dynamics \cite{gkms} with the experimental data. The histogram
in Fig.7 is the LEPTO Monte Carlo estimate of the effects of radiation
from the current jet and of hadronization.

Finally let us consider the production of dijets
close to the photon fragmentation region and with small rapidity
gap between the jets (Fig.4d).
Absence of transverse momentum ordering within the BFKL gluon ladder
modifies the theoretical expectations concerning
 production of such dijets in deep inelastic scattering when the dominant
 subprocess is that of the virtual photon-gluon fusion.
In particular one gets significant
broadening of the azimuthal distributions of the two jets which increases
with decreasing $x$  as shown in Fig.8 \cite{agkm}.
The distributions plotted in Fig.8 are normalized to a
common maximum.  The weakening of
the back-to-back correlations between the jets in photoproduction has
recently
been discussed in the ref. \cite{fr}.

Other processes which are sensitive to the small $x$ physics are the
deep inelastic diffraction and heavy quark production.

To sum up we have recalled in this talk the basic QCD expectations
concerning the small $x$ physics and discussed
their possible implications for deep
inelastic lepton scattering.  We have limited ourselves to large $Q^2$
region where perturbative QCD is expected to be applicable. Specific
problems of the low $Q^2$, low $x$ region are
discussed in the ref. \cite{bb}.

We have shown that the gluon and sea quark distributions and so
the deep inelastic structure functions should grow rapidly  with
decreasing $x$.  This rapid $x^{-\lambda}$ type of increase with $\lambda
\simeq 1/2$  is the reflection of the BFKL
Pomeron which originates from the gluon radiation.

This indefinite growth of parton distributions cannot go on forever and has
to be eventually stopped by parton screening which leads to the
parton saturation. Most probably however the saturation limit is still
irrelevant for the small $x$ region which is now being probed at HERA.

The observed
increase of $F_2(x,Q^2)$ with decreasing $x$ observed at HERA
is in agreement with the predictions
based on the BFKL equation. The explanations of this effect
within the Altarelli-Parisi evolution  equations  are also possible.

Finally we have emphasised  the role of studying the final states in
deep inelastic scattering at small $x$ and have discussed jet production,
transverse energy flow and dijet production.
The small $x$ dynamics should also show up at other semihard
processes like heavy quark production or deep inelastic diffraction.
\vskip0.5cm

{\bf Acknowledgments.}

I wish to thank Stephan Narison for  excellent organisation of the Workshop.
I thank Adrian Askew,
Barbara Bade\l{}ek, Krzysztof Golec-Biernat, Dirk Graudenz,
Alan Martin and Peter Sutton for numerous discussions and for
very enjoyable research collaboration on
problems presented in this talk.
This research has been supported in part
by the Polish Committee for Scientific Research Grant N0. 2 P302 062 04.

\vfill
\pagebreak
\phantom{of the opera}
\vskip8cm
{\small Fig. 1  The regions in the (ln$1/x$, ln$(Q^2/\Lambda^2)$)
plane where
various dynamical effects and approximations play  dominant role.}
\vskip8cm
{\small Fig. 2  Diagrammatic representation of the gluon ladder contribution
to deep inelastic scattering and of the $k_T$ factorisation formula
(\ref{ktf}).}
\vfill
\pagebreak
\twocolumn[\phantom{of the opera}
\vskip18cm
{\small Fig. 3 The measurements of the $F_2$ at HERA shown together with the
BFKL description \cite{akms1} (continuous curves)
and the MRS(A) parton analysis fit \cite{mrs3} (dashed curves).  The curves
marked as GRV at the plots for $Q^2$ = 15 GeV$^2$ and for $Q^2=$
30 GeV$^2$ correspond to the
predictions of the dynamical model
of ref. \cite{grv}.}
\vfill
\pagebreak]
\twocolumn[\phantom{in the opera}
\vskip8cm
{\small Fig. 4 Diagrammatic representation of the processes
testing the BFKL dynamics. (a) Deep inelastic scattering
with the forward jet . (b) $E_T$ flow in deep inelastic scattering.
(c) Production of jets separated by the large rapidity gap $\Delta y$.
(d) Dijet production in deep inelastic scattering.}
\vskip8cm
{\small Fig. 5 The cross section $<\sigma>$, in pb, for deep inelastic plus jet
events shown as the function of $x$ for three different bins of $Q^2$
\cite{kms2,kms3}. The continuous curves show $<\sigma>$
calculated with the inclusion of the BFKL soft gluon summation.
The corresponding $<\sigma>$ values which were calculated
neglecting the BFKL effects are shown as the dashed curves.}
\vfill
\pagebreak
\phantom{in the opera}]
\twocolumn[\phantom{in the opera}
\vskip8cm
{\small Fig. 6 (a) Diagrammatic representation of the formula (\ref{et})
showing the
gluon ladders which are resummed by the BFKL equation for $f$ and
${\cal F}_2$. (b) The representation of formula (\ref{eth}), which is
obtained by the simplification of (\ref{et}) when $x_j$ is large. }]
\phantom{in the opera}
\vskip7cm
{\small Fig. 7 The data show the $E_T$ flow accompanying deep-inelastic events
with
$x<10^{-3}$ observed by the H1 collaboration in the central region
\cite{ubass,eth1}.   The
continuous curves show the BFKL predictions
 \cite{gkms}.}
\vfill
\pagebreak
\phantom{in the opera}
\vskip8cm
{\small Fig. 8 The azimuthal distributions of the jets
for  deep inelastic dijet events \cite{agkm}.}

\end{document}